\documentclass[a4paper,12pt]{article}
\usepackage[utf8]{inputenc}
 \usepackage[margin=1.1in]{geometry}
 \usepackage{amssymb} 
\usepackage{amsmath}
\usepackage{authblk}
\usepackage{comment}
\usepackage{tikz}
\usepackage{graphicx}
\usepackage{array}
\usepackage[hyperindex,breaklinks]{hyperref}
\usepackage[numbers,sort&compress]{natbib}
\usepackage{braket}
\usepackage{slashed}

\title{\Huge Gravitational waves from \\
thermal heavy scalar dark matter}
\author{\vspace*{2cm}Parsa Ghorbani}
\affil{
\normalsize \it Physics Department, Faculty of Science, Ferdowsi University of Mashhad, Iran\\
}
\date{}

\begin{document}
\maketitle

\begin{abstract}

We propose a gauge singlet scalar with mass around $1-100$ TeV as a thermal heavy dark matter candidate along with a dilaton as a Higgs portal mediator in a dimensionless scalar extension of the Standard Model. We demonstrate analytically that such a model gives rise to very strong first-order electroweak phase transition through supercooling. We calculate the corresponding gravitational wave signals due to bubble collisions during the phase transition. The produced gravitational waves can be detected by future space-based gravitational wave detectors in the frequency range from $10^{-4}$ Hz to $0.1$ Hz. 
\end{abstract}

\newpage

\section{Introduction}
After the discovery of first spin zero scalar elementary particle at the LHC in 2012 \cite{ATLAS:2012yve,CMS:2012qbp,ATLAS:2013xga}, it has been a reasonable assumption to expect more scalars in nature. While the existence of other types of particles is possible in beyond the standard model (BSM), but an addition of even one scalar degree of freedom apart from the Higgs particle, complicates significantly the process of the electroweak symmetry breaking \cite{Ghorbani:2020xqv,Ghorbani:2021rgs}. The scalar extensions of the standard model have been always of interest to explain the theoretical and observational shortcomings of the  standard model (SM). Additional scalars may accommodate the dark matter problem \cite{McDonald:1993ex,Drozd:2011aa,Yaguna:2011qn,Gonderinger:2012rd,Campbell:2016zbp}, or to impose the electroweak symmetry breaking in the early universe be strongly first order for a successful explanation of the observed matter-anti matter asymmetry in the universe \cite{Kurup:2017dzf}, through which gravitational wave signals can be detected \cite{Vaskonen:2016yiu,Ellis:2022lft}, or to address a combination of the aforementioned problems \cite{Beniwal:2017eik,Ghorbani:2018yfr,Ghorbani:2019itr}.

Additional scalars can be employed in dimensionless extensions of the SM to address additionally the hierarchy and fine-tuning problems. There are scale invariant models that have tried to explain one or more problems such as dark matter or electroweak phase transition \cite{Khoze:2013uia,Ghorbani:2015xvz,Ghorbani:2017lyk, YaserAyazi:2018lrv,Barman:2021lot,Ahriche:2023jdq,Wong:2023qon,Liu:2024fly}. 
Although it is known that electroweak phase transition in scale invariant scalar extension of the SM is very strong \cite{Ghorbani:2017lyk}, but the possibility of detecting the gravitational waves signals produced during such phase transitions when an extra scalar plays the role of a thermal heavy dark matter has not been investigated in the literature. A nonthermal heavy dark matter scenario from first-order phase transition is studied in \cite{Giudice:2024tcp}.

The main points we will highlight in this work before going into the details are as follows:

$\bullet$ The model lies within the Gildener-Weinberg scale invariant   approach, therefore the electroweak phase transition occurs only via Coleman-Weinberg radiative corrections. 

 $\bullet$ After the symmetry breaking, there is always a heavy scalar which is stable and can be dark matter candidate. 
 
  $\bullet$ Even with very small couplings, the heavy dark matter can be produced thermally in early universe. 
  
$\bullet$ The electroweak phase transition is first-order and very strong. We show this analytically in high temperature approximation. 
  
$\bullet$  The dominant contribution in producing the gravitational waves is due to the bubble wall collisions.

$\bullet$ The gravitational waves produced by EWPT is detectable only in future GW detectors. 

The paper is organized in the following order. In section 2, we introduce the dimensionless extension of the SM, and define the free parameters and will give the effective potential after the EWSB. In section 3, we show how the EWPT in this model is strongly first-order. In section 4, we introduce the heavy dark matter candidate and calculate its relic density. We also calculate the spin-independent DM-nucleon cross section and confront the result with the direct detection bounds. In chapter 5, we calculate possible gravitational wave signals when the parameter space is restricted already by heavy dark matter. We conclude with the results in section 6.

\section{Model}
The model consists of a two gauge singlet scalars coupled to the Higgs within the scale invariant extension of the SM. Among two singlet scalars one gains mass via radiative corrections and the other is heavy according to Gildener-Weinberg \cite{Gildener:1976ih}. The heavier scalar is taken as the DM candidate while the lighter scalar, the so-called scalon field, is the SM-DM mediator.
The bare potential with adimensional couplings consisting of the Higgs doublet and two singlet scalars before the electroweak symmetry breaking (EWSB) read
\begin{equation}
V(\phi_1,\phi_2,\phi_3)=\frac{1}{4}\lambda_1 \phi_1^4+\frac{1}{4}\lambda_2 \phi_2^4+\frac{1}{4}\lambda_3 \phi_3^4+\frac{1}{2}\lambda_\text{12} \phi_1^2 \phi_2^2+\frac{1}{2}\lambda_{13} \phi_1^2 \phi_3^2+\frac{1}{2}\lambda_{23} \phi_2^2 \phi_3^2
\end{equation}
where $\phi_1$ is the neutral component of the Higgs doublet
$H^\dagger = (0~\phi_1)/\sqrt{2}$ in unitary gauge, and $\phi_2$ and $\phi_3$ are real singlet scalars. 

The scale symmetry breaking which subsequently leads to electroweak symmetry breaking takes place through radiative corrections. According to Gildener-Weinberg approach there exists a flat direction in the field space along which the potential and its minimum is vanishing. In our model, the flat direction with three scalar field is three-dimensional $(n_1,n_2,n_3)$. As the third scalar, i.e. $\phi_3$ is the DM candidate here, it does not acquire a non-zero VEV, therefore we  set $n_3=0$ and the flat direction will be in $(\phi_1, \phi_2)$ space. 
\begin{equation}
\phi_1=n_1 \phi\equiv  \phi \sin\theta, ~~~~~~~~~~\phi_2=n_2 \phi \equiv  \phi \cos\theta,
\end{equation}
where $n_1^2+n_2^2=1$ and $\phi^2=\phi_1^2+\phi_2^2 $ is the radial field. The flat direction is given by
\begin{equation}\label{flatc}
\sin^2{\theta} \equiv n_1^2=\frac{\lambda_{12}}{\lambda_{12}- \lambda_1}, ~~~~~~~~\cos^2{\theta} \equiv n_2^2=\frac{\lambda_{1}}{\lambda_{1}- \lambda_{12}}.
\end{equation}
The Coleman-Weinberg one-loop effective potential along the flat direction is given by 
\begin{equation}\label{veff}
V_\text{eff}^\text{1-loop}(\phi)= B \phi^4\left( \log{\frac{\phi}{\braket{\phi}}-\frac{1}{4}}\right)
\end{equation}
where $\braket{\phi}$ is the VEV of the radial field and $B$ is a dimensionless coupling given in general by
\begin{equation}
B=\frac{1}{64\pi^2\braket{\phi}^4}\left(\text{Tr}M_S^4+3\text{Tr}M_V^4-4\text{Tr}M_F^4 \right)
\end{equation}
in which $M_S$, $M_V$ and $M_F$ denote respectively the mass matrices for scalars, vectors and fermions in the model. In particular, for the  model we are considering
\begin{equation}
B=\frac{1}{64\pi^2\braket{\phi}^4}\left(m_{DM}^4+m_H^4+6m_W^4+3m_Z^4-12m_t^4  \right)
\end{equation}
where $m_\text{DM}\equiv m_{\phi_3}$. The effective potential in Eq. (\ref{veff}), in addition to having a minimum at zero it  develops a deeper minimum for a non-zero VEV, so that the symmetry is radiatively broken at one-loop level. Through the VEV of the radial field, $\braket{\phi}$, scalars $\phi_1$ and $\phi_2$ take non-zero VEVs. 
Subsequently, there is a mixing among scalars $\phi_1$ and $\phi_2$ and their mass matrix will not be diagonalized. Rotating the $(\phi_1, \phi_2)$ into a new basis $(h,s)$ defined as  
\begin{equation}
\begin{pmatrix}
h \\
s \\
\end{pmatrix}=\begin{pmatrix}
\cos{\theta} & \sin{\theta} \\
-\sin{\theta} & \cos{\theta}
\end{pmatrix}
\begin{pmatrix}
\phi_1 \\
\phi_2 
\end{pmatrix}
\end{equation}
will diagonalize the mass matrix with mass eigenvalues
\begin{subequations}
\begin{align}
&m_H^2=-2\lambda_{12} \braket{\phi}^2,\\
&m_\text{DM}^2=  (\lambda_{13}\sin^2{\theta}+ \lambda_{23}\cos^2{\theta})\braket{\phi}^2,\\
& m_s=0.
\end{align}
\end{subequations}
As anticipated in GW approach one of the scalars (here scalar $s$) is massless which is known as {\it scalon}. 
The dynamical symmetry breaking gives a small mass to the classically massless scalon $s$
\begin{equation}
\delta m^2_\text{s}=8B\braket{\phi}^2=\frac{1}{8\pi^2 \braket{\phi}^2}(m^4_H+6 m^4_W+3 m^4_Z-12 m^4_t+m^4_\text{DM}).
\end{equation}
 Considering the latest experimental values for heavy SM particle masses being $m_H=125.12$ GeV, $m_Z=91.19$ GeV, $m_W=80.38$ GeV, $m_t=172.76$ GeV, and the fact that the scalon mass must be  positive, i.e. $\delta m^2_\text{s}>0$, the DM mass is bounded from below: $m_\text{DM}>316.12$ GeV. In order to get small scalon mass, its VEV must be considerably large; for instance if $v_s=1000$ GeV then $m_\text{s}\simeq 10$ GeV. 
The effective potential after the electroweak symmetry breaking read
\begin{equation}
 \label{vtreff}
\begin{split}
V_\text{tr}(h,s,\phi_3)=&-\lambda_{12} \Big[  \braket{\phi}^2 h^2 + 2\braket{\phi} h^3 \cot{(2\theta)} +h^4 \cot^2{(2\theta)}  \\
&+h^2 s^2 
+2 \braket{\phi}h^2 s +2 h^3 s \cot{(2\theta)} \Big] \\
&+\frac{1 }{2} \Big[ \lambda_{13} \sin^2(\theta)+ \lambda_{23} \cos^2{(\theta)}\Big] \Big[ s^2 \phi_3^2   +  \braket{\phi}^2\phi_3^2 +2  \braket{\phi} s \phi_3^2\Big] \\
&+\frac{1 }{2} \Big[ \lambda_{13} \cos^2{(\theta)}+ \lambda_{23} \sin^2{(\theta)}\Big] h^2 \phi_3^2   \\
&+\frac{1}{2}(\lambda_{13}-\lambda_{23})  \braket{\phi} h \phi_3^2 \sin{(2\theta)}
+\frac{1}{4}\lambda_3 \phi_3^4
\end{split}
\end{equation}
where $\theta$ is the mixing angle. 

\section{Heavy Scalar  Dark Matter}
From the LHC Higgs phenomenology the mixing angle is constrained as $\cos{\theta}>0.85$ or $-0.555<\theta<0.555$. The Higgs field VEV is fixed $v_h=246$ GeV, and the singlet scalar VEV is determined through the scale symmetry breaking scale $\braket{\phi}$ by $\braket{\phi}^2=v_h^2+v_s^2$. Taking into account that $\tan{\theta}=v_h/v_s$ and choosing the scale $\braket{\phi}$ as a free parameter, the mixing angle is fixed by $\cot^2{\theta}=\braket{\phi}^2/v_h^2-1$. The larger the scale of $\braket{\phi}$, the smaller the mixing angle will be; the minimum value is $\braket{\phi}_\text{min}\simeq 466$ GeV for $\theta \simeq 0.555$. The coupling $\lambda_{12}$ is determined by $m_H^2=-2\lambda_{12}\braket{\phi}^2$ with $m_H\simeq 125$ GeV. The coupling $\lambda_1$ is not free as it is fixed from Eq. (\ref{flatc}). Similarly for the coupling $\lambda_2$ which is obtained from the flat direction condition $\lambda_2=\lambda_{12}^2/\lambda_1$. Therefore, the space of free parameters is as restricted as $\{\braket{\phi},\lambda_{13},\lambda_{23}, \lambda_3 \}$. 
As pointed out before the dark matter candidate is the field $\phi_3$ with vanishing VEV which makes it a stable particle. Its relic density is obtained by the Boltzmann equation for thermal evolution of dark matter field number density $n_\text{DM}$, 
\begin{equation}
\frac{dn_\text{DM}}{dt}+3Hn_\text{DM}=-\braket{\sigma_\text{ann}v_\text{rel}}\big[ n_\text{DM}^2 -{n_\text{DM}^\text{eq}}^2\big]
\end{equation}
in which $n_\text{DM}^\text{eq}$ stands for the DM number density in plasma thermal equilibrium in the early universe, $H$ is the Hubble expansion rate, $v_\text{rel}$ is the dark matter relative velocity and $\sigma_\text{ann}$ is the dark matter annihilation cross section. 
\begin{figure}\label{dm-nuc}
\begin{center}
\includegraphics[scale=0.7]{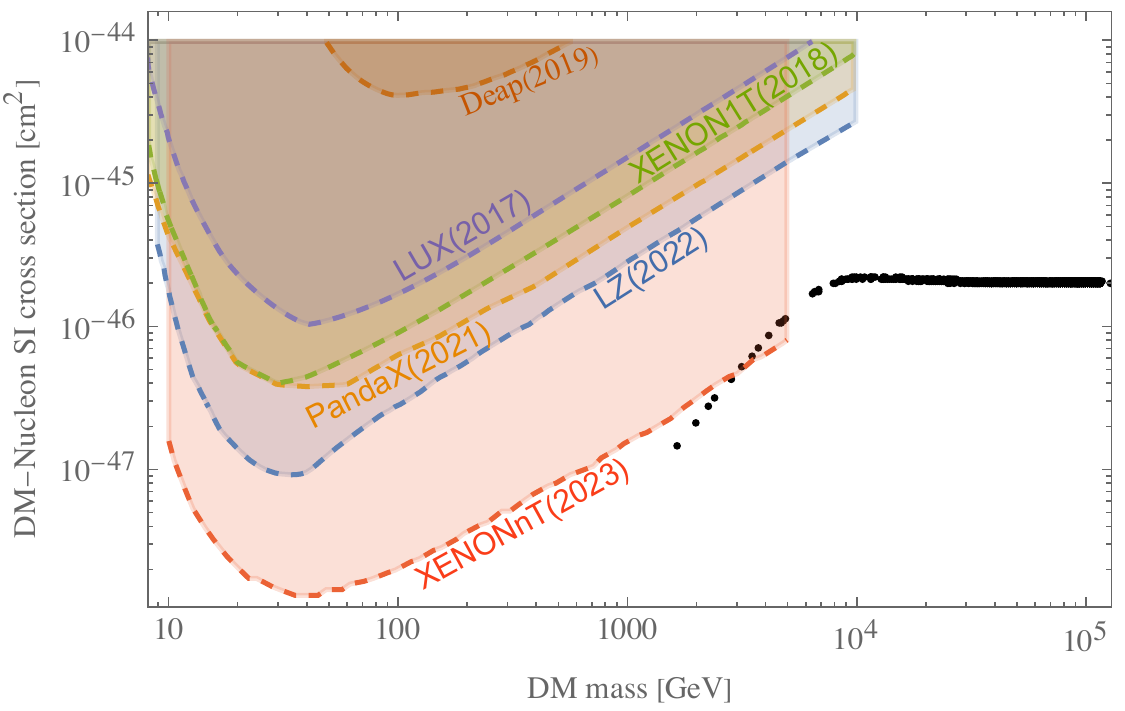}
\end{center}
\caption{The DM-nucleon cross section is shown. Almost all the parameter space for DM masses heavier than $1.5$ TeV is evaded from direct detection constraints. }
\end{figure}

\begin{figure}[!t]
\begin{center}
\includegraphics[scale=0.5]{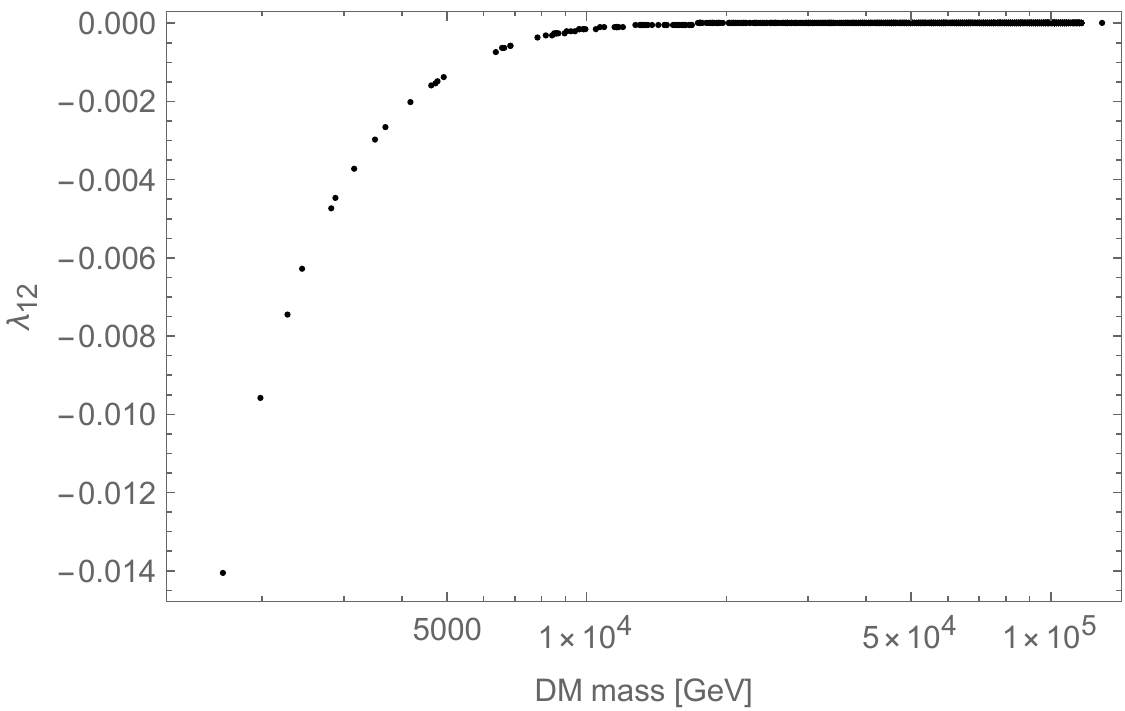}
\includegraphics[scale=0.5]{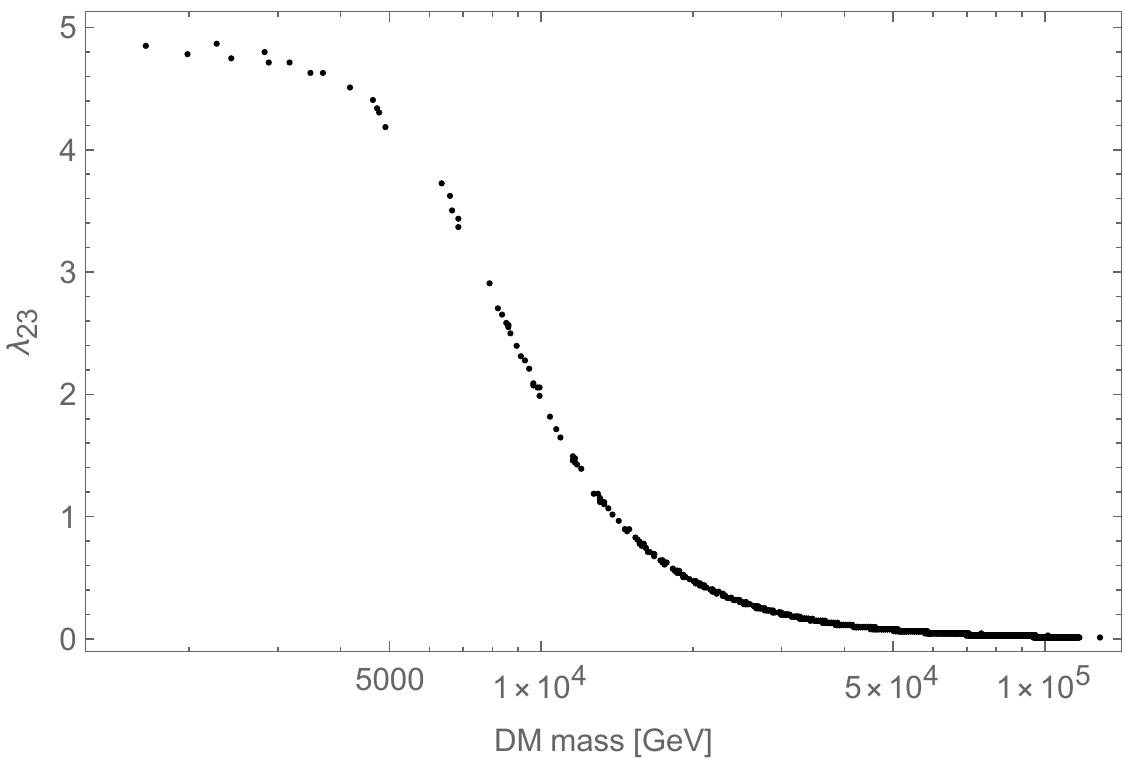}
\includegraphics[scale=0.5]{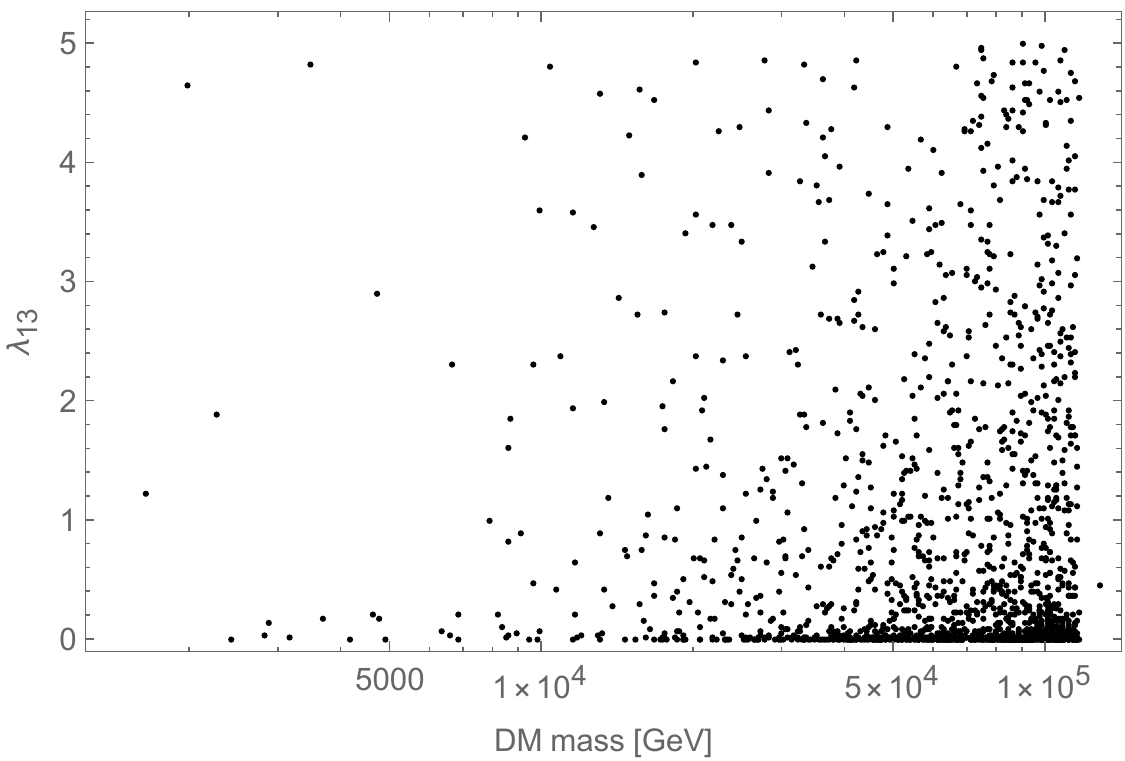}
\end{center}
\caption{Shown are the DM mass against the Higgs portal coupling $\lambda_{12}$, and the coupling $\lambda_{13}$, and $\lambda_{23}$ in the model. The DM relic abundance is restricted to be the observational value $\Omega_\text{DM}h^2=0.12$. Smaller DM mass requires larger Higgs portal coupling.}
\label{relic123}
\end{figure}

We have used the package {\tt MicrOMEGAs6.0} to numerically calculate the relic abundance of the heavy dark matter. The model accommodates the  DM relic density $\Omega_\text{DM}h^2 \simeq 0.12$ observed by WMAP and Planck \cite{Hinshaw:2012aka, Planck:2018vyg}. 
The results in Figs. \ref{relic123} show that respecting the observed relic density value, the lighter DM requires larger Higss portal coupling $\lambda_{12}$. Inversely, the larger DM mass requires larger coupling $\lambda_{23}$. The DM mass predicted in this model is above $1.5$ TeV. 

The parameter space in the model is strongly constrained by DM-nucleon cross section the direct detection experiments \cite{Aprile:2017iyp,XENON:2018voc,XENON:2023cxc,PandaX-4T:2021bab,LZ:2022lsv,LUX:2016ggv,DEAP:2019yzn}. The effective DM-nucleon interaction Lagrangian is 
\begin{equation}
\mathcal{L}_\text{eff}=\alpha_q \phi_3 \phi_3 \bar q q
\end{equation}
where $\alpha_q$ is the DM-quark coupling. This leads to spin-independent (SI) DM-Nucleon scattering cross section 
\begin{equation}
\sigma_\text{SI}^\text{N}=\frac{\alpha_N^2 \mu_N^2}{\pi m_\text{DM}^2}
\end{equation}
with $\mu_N$ being the reduced DM-Nucleon mass and $\alpha_N$ a coefficient depending  on nucleon form factors \cite{Ghorbani:2014qpa}.  We have used the {\tt MicrOMEGAs6.0} package to calculate the DM-nucleon scattering cross section. The result is shown in Fig. \ref{dm-nuc}. Although the parameter space is already restricted from the DM relic abundance, we see in Fig. \ref{dm-nuc} that except from a small region of the parameter space, the model is evaded from the direct detection constraints. 
\section{Electroweak Phase Transition}
We assume that the scale symmetry is broken at high enough temperatures respect to the masses in the model so that we can use the high-temperature expansion. \footnote{If $m/T \lesssim 1.6$, the high-temperature approximation agrees better than $5\%$ with the exact thermal potential. For large values of $m/T$, still the high-temperature approximation is valid up to $10\%$ due to suppressed Boltzmann contribution (see appendix B in \cite{Anderson:1991zb}).}
The thermal effective potential in the high-temperature approximation is given by \cite{Marzola:2017jzl}
\begin{equation}\label{veff}
V_\text{eff}(T)=V+V_T=B \phi^4\left(  \log{\left(\frac{\phi}{\Lambda}\right)} -\frac{1}{4}\right)+C \phi^2 T^2
\end{equation}
with \cite{Sannino:2015wka}
\begin{subequations}
\begin{align}
&B= \frac{1}{64\pi^2}\sum_k  g_k  W_k^4(\theta)\\
&C= \frac{1}{12}\sum_k c_k g_k W_k^2(\theta),
\end{align}
\end{subequations}
where $\theta$ is the angle which rotates $(\phi_1,\phi_2)$ along the flat directions i.e. $(\phi_1\sin\theta,\phi_2\cos\theta)$, $g_k$ is the number of degrees of freedom for each particle, $c_k=1$ ($c_k=-1/2$) for bosons (fermions), and $W_k$ is defined through $M_k(\theta ,\phi)= W_k(\theta) \phi$. For the current model we have 
\begin{subequations}
\begin{align}
 &W^2_H=-2\lambda_{12}\\
 &W^2_\text{DM}=\lambda_{13}\sin^2{\theta}+\lambda_{23}\cos^2{\theta}\\
 &W^2_{W^\pm}=\frac{1}{4}g^2 \sin^2{\theta}\\
 &W^2_Z=\frac{1}{4}(g^2+g'^2) \sin^2{\theta}\\
 & W^2_t=\frac{1}{2}y^2_t \sin^2{\theta}
 \end{align}
\end{subequations}
where $g\simeq 0.648$, $g'\simeq 0.359 $ and $y_t\simeq 0.951$ are $SU(2)$ gauge coupling, $U(1)_Y$) gauge coupling,  and top quark Yukawa coupling at top mass scale , respectively. The parameters $B$ and $C$ are
\begin{subequations}\label{bc}
\begin{align}
&B=\frac{1}{64\pi^2}\left(W_H^4+6W_{W^\pm}^4+3W_Z^4-12W_t^4+W_\text{DM}^4 \right)\\
&C=\frac{1}{12} \left(W_H^2+6W_{W^\pm}^2+3W_Z^2+6W_t^2+W_\text{DM}^2 \right)
\end{align}
\end{subequations}
The effective potential in Eq. (\ref{veff}) has three extrema being $\phi_\text{min} \equiv \phi_\text{sym}=0$, 
\begin{equation}
\phi_\text{max}=\Lambda~ e^{\frac{1}{2}W_{-1}\left(-CT^2/B\Lambda^2 \right)},
\end{equation} 
and 
\begin{equation}\label{phibrk}
\phi_\text{min} \equiv \phi_\text{brk}(T)=\Lambda ~e^{\frac{1}{2}W_0\left(-CT^2/B\Lambda^2 \right)},
\end{equation}
%$v_\text{brk}(T)$
where $W_0$ and $W_{-1}$ denote respectively the principal branch and the lower real branch of the {\it Lambert $W$ function}. As the argument of the $W$  function in Eq. (\ref{phibrk}) is negative, the following condition must be held for the function to be single valued 
\begin{equation}
-CT^2/B\Lambda^2 \geq -1/e,
\end{equation}
that is an upper limit for temperature $T^2\leq B\Lambda^2/Ce$, to have a non-zero VEV for the radial field $\phi$. Above this temperature limit the effective potential is in its symmetric phase with vanishing VEV. 

The critical temperature is defined as  the temperature at which the thermal effective potential acquires two degenerate minima, one in  symmetric phase and the other in broken phase. By requiring the condition $V_\text{eff}(v_\text{brk})=V_\text{eff}(v_\text{sym})=0$ we get
\begin{equation}
T_c= \sqrt{\frac{B}{2C}} \Lambda  e^{-1/4}.
\end{equation}
All phase transitions in scale invariant models are of first order type \cite{Ghorbani:2017lyk,Marzola:2017jzl}. The phase transition will be strong if $v_\text{brk}(T_c)/T_c>1$ or equivalently 
\begin{equation}
\sqrt{\frac{2C}{B}}>1.
\end{equation} 
which is always true because from Eq. (\ref{bc}),  $c>b$.

\section{Gravitational Waves}

\begin{table}[t!]
    \centering
    \resizebox{\textwidth}{!}{
    \begin{tabular}{|c|c|c|c|c|c|c|c|}
        \hline
        $ M_\text{DM}$ & $ T_c$  & $T_n $ & $\alpha$ & $\beta/H_* $ & $\lambda_{12}$ & $\lambda_{13}$ & $\lambda_{23}$ \\
        \hline \hline
        $346.25$ GeV & $114.73$ GeV & $65$ GeV & $3.28$ & $227$ & $-1.4\times 10^{-2}$ & $1.22$ & $4.85$ \\
        \hline
         $1.69$ TeV & $911$ GeV &$217$ GeV & $103$ & $243$ & $-7\times 10^{-5} $  & $2.37\times 10^{-2}$ & $1.4$ \\
        \hline
         $26.37$ TeV & $2$ TeV & $149$ GeV & $1.1\times 10^4$ & $260$ & $-3.08\times 10^{-6}$ & $2.48\times 10^{-6}$ & $0.27$ \\
        \hline
       $36.93$ TeV & $2.8$ TeV& $117$ GeV & $1.1\times 10^5$ & $252$ & $-2.27\times 10^{-7}$ & $1.32\times 10^{-1}$ & $1.22\times 10^{-1}$ \\
 
        \hline
        $81.85$ TeV & $6.21$ TeV & $47$ GeV & $1.08\times 10^8$  & $1190$ &$-3.11\times 10^{-8}$  &$1.73$ & $2.67 \times 10^{-2}$ \\
            \hline
        $116.64$ TeV & $8.86$ TeV & $177$ GeV & $2.14\times 10^6$  & $4521$ &$-7\times 10^{-9}$  &$4.53$ & $1.36 \times 10^{-2}$ \\
        \hline
    \end{tabular}
    }
    \caption{Shown are benchmarks for strongly first-order electroweak phase transition and gravitational wave parameters. For the four last benchmarks, the observed DM relic density cannot be fully accounted by the model due to DM dilution following the phase transition.}
    \label{gwtabel}
\end{table}

Gravitational waves might stem from strong first-order phase transition in early universe in different ways (see e.g. \cite{Caprini:2015zlo}): 1) the bubble walls collisions and shocks in the plasma, $\Omega_\phi h^2$, which is the contribution of the scalar field $\phi$ using a technique known as 'envelope approximation', 2) contributions from the sound waves, $\Omega_\text{sw} h^2$, produced by the bubble wall collision and 3) from the magnetohydrodynamic (MHD) turbulence in the plasma $\Omega_\text{tur}$.  The total stochastic GW background is approximately a linear combination of all contributions,
\begin{equation}
\Omega_\text{GW}h^2 \simeq \Omega_\phi h^2+\Omega_\text{sw} h^2+\Omega_\text{tur}
\end{equation}

Two key parameters are used in GW contributions; one is the ratio of the vacuum energy density released during the phase transition to the radiation energy density \cite{Espinosa:2010hh},
\begin{equation}
\alpha(T)=\frac{\Delta\epsilon(T)}{\rho_\text{rad}(T)}
\end{equation}
where 
\begin{equation}
\rho_\text{rad}=\pi^2 g_* T^4/30
\end{equation}
is the radiation energy density, and 
\begin{equation}
\Delta\epsilon(T)=\epsilon(\phi_\text{brk},T)-\epsilon(\phi_\text{sym},T)
\end{equation}
with
\begin{equation}
\epsilon(\phi,T)=-V_\text{eff}(\phi,T)+\frac{T}{4} ~ \frac{d V_\text{eff}(\phi,T)}{d T} 
\end{equation}
Noting that from Eq. (\ref{veff}),  $V_\text{eff}(\phi_\text{sym}=0,T)=0$ and $dV_\text{eff}(\phi_\text{sym}=0,T)/dT=0$, that results in $\epsilon(\phi_\text{sym}=0,T)=0$, therefore we have 
\begin{equation}
\Delta\epsilon(T)=\frac{1}{4}b\Lambda^4 e^{2 W_0(-cT^2/b\Lambda^2)}
\end{equation}
and 
\begin{equation}
\alpha(T)=\frac{15B}{2\pi^2 g_*}\left(\frac{\Lambda}{T}\right)^4 e^{2 W_0(-CT^2/B\Lambda^2)}
\end{equation}
The other parameter is the ratio of the inverse time duration of the phase transition, $\beta$, to the Hubble parameter $H_*$ at  temperature $T_*$,
\begin{equation}
\frac{\beta}{H_*}= T \frac{d}{dT}\left( \frac{S_3(T)}{T} \right)\bigg\vert_{T_*}.
\end{equation}

The temperature at which the gravitational waves are produced is denoted usually by $T_*$. We assume the reheating is negligible, so that the temperature at which the gravitational waves are produced is almost the nucleation temperature i.e. $T_*\simeq T_n$. 
The bubble nucleation rate per unit volume is given by \cite{Linde:1981zj}
\begin{equation}
\Gamma(T) \simeq T^4 \left( \frac{S_3(\phi,T)}{2\pi T}\right)^{3/2}  e^{S_3(\phi,T)/T}
\end{equation}
where $S_3(\phi,T)$ is the action for $O(3)$ symmetric bubble
\begin{equation}
S_3=4\pi \int{dr \, r^2 \left[\frac{1}{2} \left(\frac{d\phi}{dr} \right)^2+V_\text{eff}(\phi,T) \right]}.
\end{equation}
which is to be minimized by the  $\phi$ (radial field) profile from $\braket{\phi}=0$ to $\braket{\phi} \neq 0$. Here, $V_\text{eff}(\phi,T)$ is the one-loop thermal effective potential in Eq. (\ref{veff}). The configuration which minimizes $S_3$ is the solution to 
 \begin{equation}
\frac{d^2\phi}{dr^2}+\frac{2}{r}\frac{d\phi}{dr}=\frac{dV_\text{eff}}{d\phi},
\end{equation}
with boundary conditions $\phi=0$ when $r\to \infty$ and $d\phi/dr=0$ at $r=0$. 

For very strong electroweak phase transition the dominant contribution source of gravitational waves is due to the scalar field which is given by the envelope approximation \cite{Kosowsky:1991ua,Kosowsky:1992rz,Kosowsky:1992vn,Kamionkowski:1993fg}
\begin{equation}
\Omega_\text{GW} h^2=\left( \frac{H_*}{\beta} \right)^2 \left( \frac{100}{g_*}\right) \frac{4.9\times 10^{-6} (f/f_\text{env})^{2.8}}{1+2.8(f/f_\text{env})^{3.8}}
\end{equation}
where $H_*$ is the Hubble parameter at the reheating temperature $T_*\simeq T_n$, $f_\text{env}$ is the peak frequency of the spectrum in envelope approximation,
\begin{equation}
\frac{f_\text{env}}{\text{Hz}}=3.5\times 10^{-6} \left(\frac{\beta}{H_*} \right) \left( \frac{g_*}{100}\right)^{1/6} \left( \frac{T_*}{100\,\text{GeV}}\right) 
\end{equation}

%The GW signal frequency and plot data was taken from \cite{Athron:2023xlk}.

\begin{figure}[!t]
\begin{center}
\includegraphics[width=\textwidth]{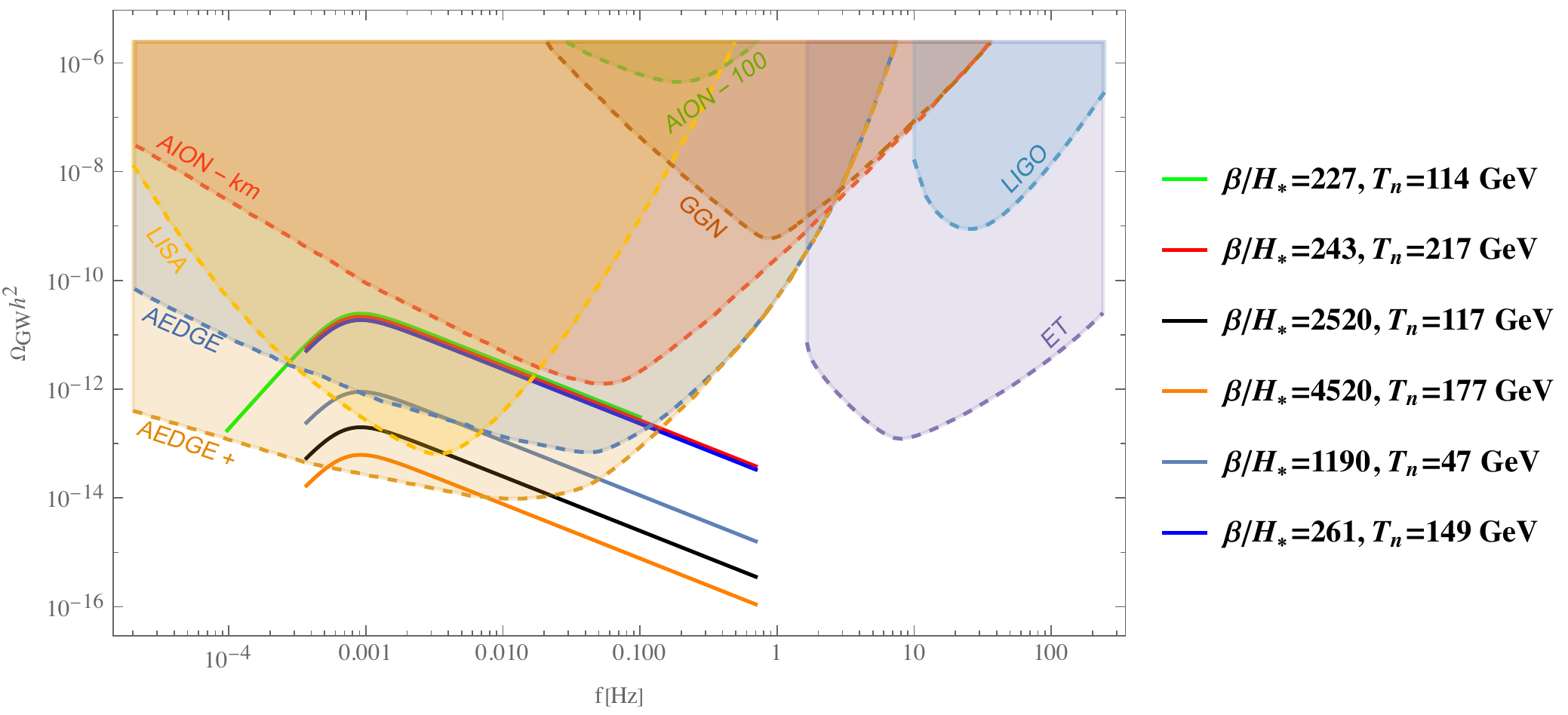}
\end{center}
\caption{Shown is the gravitational wave signals predicted from the benchmarks in table 1 to be detected by space-based GW detectors.}
\label{gwsignal}
\end{figure}

In table \ref{gwtabel}, we have chosen from  lightest to heaviest DM mass possible with corresponding parameters which give rise to the correct observed DM relic abundance. For the DM mass $m_\text{DM}\sim 350$ GeV the coupling $\lambda_{12}$ is of order $10^{-2}$ while for the heaviest DM mass in the table, $m_\text{DM}\sim 100$ TeV the coupling $\lambda_{12}$ becomes as small as $10^{-9}$. For the parameter $\alpha$ that is opposite; the heavier the DM mass is, the larger parameter $\alpha$ becomes, i.e. from $\alpha \sim 3$  to $\alpha \sim 10^6$ . For large $\alpha$, due to substantial amount of latent heat and therefore a large entropy injection into the SM sector, the relic density will be diluted \cite{Hambye:2018qjv}. Taking into account this effect, the four last benchmarks in Table 1, cannot account for all observed DM relic abundance.
Another point in table \ref{gwtabel} is that for heavy DM there is a suppercooling in electroweak phase transitions, which is of course expected from  the large values for the $\alpha$ parameter.  Although the critical temperature lies around TeV scale, the nucleation temperature is always around the electroweak scale. 
For the benchmarks given in table \ref{gwtabel} we have plotted the gravitational waves signal produced during the bubble wall collisions in terms of the nucleation temperature $T_n$ and the ratio $\beta/H_*$. The result is shown in figure \ref{gwsignal}. We see that the for heavy dark matter candidates in the current model, the GW signals can be detected only by AEDGE+, AEDGE and LISA with GW frequency from around $10^{-4}$ Hz to $0.2$ Hz. This is compatible with the results in \cite{Khoze:2022nyt}.

\section{Conclusion}
%https://arxiv.org/pdf/1704.01034.pdf:
%Phase transition and gravitational wave phenomenology of scalar conformal extensions of the Standard Model; 
%Luca Marzola, Antonio Racioppi, and Ville Vaskonen

We have proposed a dimensionless extension of the standard model with two extra gauge singlet scalars coupled to the standard model through a Higss portal. In scale invariant models, the electroweak symmetry breaking occurs merely via radiation corrections a la Coleman-Weinberg. If one of the scalars does not get a non-zero VEV, it can represent the particle of dark matter. 
We have examined the current model when the scale of scale symmetry breaking is above TeV. At these high scales, the dark matter becomes heavy. A characteristic of this model is that even for large dark matter masses the dimensionless couplings remain very small while the direct detection constrains are evaded easily.  The dark matter mass that the model predicts is above $1.5$ TeV. 
We have also shown analytically that the electroweak phase transition in this model is first-order and very strong. For a set of benchmarks that the relic density of dark matter is that of the observed value by WMAP and Planck, we have calculated the gravitational waves signals due to the bubble collisions during the strong electroweak phase transitions. The result is that the GW signals can be detected by future space-based gravitational wave detector AEDGE+, AEDGE and LISA in the frequency range from $10^{-4}$ Hz to $0.1$ Hz. 

\section*{Acknowledgement}
I would like to acknowledge support from the ICTP through the Associates Programme (2024-2029). I would also like to thank the TH department at CERN  for the hospitality and support during my visit in August 2024.  
\bibliography{ref.bib}
\bibliographystyle{unsrt}
\end{document}